\documentclass[conference, letterpaper]{IEEEtran}
\ifCLASSINFOpdf
\else
\fi
\ifCLASSINFOpdf
   \usepackage[pdftex]{graphicx}
\else
\fi

\usepackage[cmex10]{amsmath}
\usepackage{color}
\usepackage{fancyhdr}
\usepackage[caption=false,font=footnotesize]{subfig}

\renewcommand{\thispagestyle}[2]{}

\fancypagestyle{plain}{
        \fancyhead{}
        \fancyhead[C]{first page center header}
        \fancyfoot{}
        \fancyfoot[C]{first page center footer}
}
\newcommand{\subsubsubsection}[1]{\paragraph{#1}\mbox{}}
\usepackage{hyperref}
\usepackage{algorithm}
\usepackage{algpseudocode}
\usepackage{fancyhdr}
\usepackage{booktabs}
\usepackage{tabularx}
\usepackage{flushend}

\begin{document}

%
\title{Comparing AI Algorithms for Optimizing Elliptic Curve Cryptography Parameters in E-Commerce Integrations: A Pre-Quantum Analysis}




%
\author{
  \IEEEauthorblockN{
    Felipe Tellez,
    Jorge Ortíz
  }
  \IEEEauthorblockA{
    Department of Systems and Industrial Engineering, National University of Colombia, Bogotá, Colombia 111321
  }
}


\maketitle

\begin{abstract}
This paper presents a comparative analysis between the Genetic Algorithm (GA) and Particle Swarm Optimization (PSO), two vital artificial intelligence algorithms, focusing on optimizing Elliptic Curve Cryptography (ECC) parameters. These encompass the elliptic curve coefficients, prime number, generator point, group order, and cofactor. The study provides insights into which of the bio-inspired algorithms yields better optimization results for ECC configurations, examining performances under the same fitness function. This function incorporates methods to ensure robust ECC parameters, including assessing for singular or anomalous curves and applying Pollard's rho attack and Hasse's theorem for optimization precision. The optimized parameters generated by GA and PSO are tested in a simulated e-commerce environment, contrasting with well-known curves like secp256k1 during the transmission of order messages using Elliptic Curve-Diffie Hellman (ECDH) and Hash-based Message Authentication Code (HMAC). Focusing on traditional computing in the pre-quantum era, this research highlights the efficacy of GA and PSO in ECC optimization, with implications for enhancing cybersecurity in third-party e-commerce integrations. We recommend the immediate consideration of these findings before quantum computing's widespread adoption.
\end{abstract}



%
\IEEEpeerreviewmaketitle

\section{Introduction}
This paper explores the field of Elliptic Curve Cryptography (ECC), a form of public key cryptography that uses the mathematics of elliptic curves to secure transactions, specifically focusing on its application within e-commerce transactions executed through third-party integrations during a pre-quantum computing era. The aim is to identify the most efficient and effective Artificial Intelligence (AI) algorithm for optimizing parameters essential to ECC's successful operation within this context. The two leading algorithms being examined are Genetic Algorithms (GA), and Particle Swarm Optimization (PSO).

\subsection{Background}
Elliptic Curve Cryptography (ECC), an important and widely used form of public-key cryptography [1][2], provides enhanced security with shorter key lengths [3], making it ideal for resource-constrained environments like e-commerce platforms [34]. Effective ECC operation hinges on the careful selection of parameters such as curve coefficients, base point, prime modulus, and others [4] – [11]. Optimizing these parameters can enhance ECC's security and efficiency, which is crucial in e-commerce transactions.

Artificial Intelligence (AI) has shown tremendous potential in this parameter optimization [12] - [26]. Notably, two AI algorithms, -GA, and PSO- stand out [12][15]. They represent different subsets and categories of AI algorithms: GA, an Evolutionary Algorithm [18], and PSO, a Swarm Intelligence [19], belong to Population-based Optimization. These algorithms are recognized for their problem-solving and optimization capabilities.

E-commerce transactions often involve integrations with third-party solutions such as Enterprise Resource Planning (ERP) systems, Customer Relationship Management (CRM) systems, payment gateways, data analytics solutions, among others [27][28]. These transactions need to be securely encrypted, making ECC an excellent choice, especially with optimized parameters.

\subsection{Objective}
The objective of this paper is to compare the efficacy of GA and PSO in optimizing ECC parameters for e-commerce transactions involving third-party integrations within binary computing. Specifically, the comparison aims to determine which AI algorithm is most efficient in terms of its behavior during the optimization process, and which AI algorithm is most effective in terms of the quality of the results it produces.

\subsection{Scope and Limitations}
While ECC can be used in contexts other than e-commerce third-party integrations, such as Web browsers, customer authentication, administrative user authentication, and database persistence, the focus of this article is on transaction integrations with third parties. Interactions between e-commerce systems and backend or third-party solutions such as ERPs, CRMs, payment gateways, and billers are all part of it.

This research uses a simulated e-commerce environment that incorporates a business process that entails creating orders in an emulated ERP. The information for these orders is sent from the e-commerce solution through third-party integrations using web services to imitate real-world conditions. The research relies on API simulations for outbound connections utilizing datasets relevant to these types of scenarios rather than a whole e-commerce solution.

The research also limits its scope to the pre-quantum computing era. Although quantum computing promises significant advances, its implications for ECC and AI algorithms are beyond this study's scope. The research also excludes other AI techniques like Simulated Annealing, Ant Colony Optimization or Artificial Neural Networks, despite their applicability to ECC optimization, to keep the research focused.

\subsection{Structure and Contributions}
The rest of this paper is organized as follows: Section II reviews related work on ECC optimization, AI techniques, e-commerce integrations, and pre-quantum developments. Section III details materials and methods, including ECC parameters and optimization criteria. Section IV describes the simulation environment design and implementation. Section V presents results and analysis, covering AI algorithm execution, e-commerce simulation, and comparison based on ECC criteria. Section VI discusses future improvements and limitations, including parameter tuning, parallelization, hybrid algorithms, alternative AI techniques, fitness function improvements, diverse cryptographic threats, and quantum computing implications. Section VII concludes with future work and recommendations. Unique contributions include a detailed comparison of GA and PSO for ECC optimization in e-commerce, a novel fitness function, and an evaluation framework for pre-quantum computing.

\section{Literature Review}

\subsection{ECC Parameter Optimization}
The complexity of Elliptic Curve Cryptography (ECC) optimization is central to research on ECC systems' effectiveness, security, and efficiency. Koblitz [1] and Miller [2] independently introduced Elliptic Curves in public-key cryptography, stressing the careful choice of parameters for enhanced security. Washington [3] emphasized optimizing ECC parameters such as curve coefficients, base point, prime modulus, and key sizes, all influencing ECC's performance.

Lenstra and Verheul [4] advocated ECC's use in cryptographic systems, with a focus on proper parameter selection, especially prime modulus. Blake, Seroussi, and Smart [5] delved into the details of ECC parameter selection, highlighting the selection of the base point, curve coefficients, and prime modulus.

A significant aspect of ECC optimization is speeding up point multiplication, a core ECC operation. Hankerson, Menezes, and Vanstone [6] outlined strategies for this focusing on ECC's computational aspects. Other researchers have examined ECC optimization in MANET and Sensor networks, where hardware plays a significant role. In [7] we can find a state of the art of ECC optimizations in these types of scenarios.

In general, the literature review emphasizes ECC parameter optimization's importance and complexity, covering domains such as elliptic curves' mathematical foundations, intricate parameter selection, implementation optimizations, and AI algorithms for multi-objective optimization [8] – [11]. This substantial knowledge forms the foundation of this study.

\subsection{AI Techniques for ECC Parameter Optimization}
Artificial Intelligence (AI) exhibits vast potential in ECC parameter optimization, with prominent techniques like Genetic Algorithms (GA) [12] - [14] and Particle Swarm Optimization (PSO) [15] - [17]. These methods contribute distinctive strengths to ECC optimization.

GA, inspired by biological evolution, is known for its effectiveness in exploring complex search spaces, particularly in seeking optimal solutions within intricate landscapes like ECC parameter optimization [18]. On the other hand, PSO, modeled after the social behaviors of birds and fishes, is celebrated for its simple implementation and intrinsic ability to avoid local optima [19]. These methods, by emulating natural processes, present unique solutions to ECC's challenges, highlighting the connection between nature's complexity and technological innovation.

Besides these techniques, others like Simulated Annealing (SA) [20], an Stochastic Optimization inspired by the annealing process in metallurgy, is known for adaptability and robustness in solving optimization issues [21], including ECC. Evolutionary Algorithms (EA), similar to GA, involve mechanisms like reproduction and mutation, showing promise in optimizing Elliptic Curves [22][23]. Machine Learning (ML), where algorithms evolve through data usage, has been applied to Elliptic Curve factorization problems [24]. Also, Tabu Search has been used to enhance ECC operations and multimedia encryption [25] [26].

Each of these named AI methods offers unique benefits in the optimization of ECC parameters; nevertheless, as we mentioned in the introduction section, the focus of our analysis will be on GA, and PSO.

\subsection{E-commerce and Third-party Integrations}
The growth of e-commerce has fostered an interconnected technological network involving various third-party entities such as payment gateways, ERP systems, CRM systems, billers, web services, and custom solutions [27][28]. They exchange vital operational data, including inventory status, billing data, orders information and more.

\subsubsection{Types of E-commerce Integrations}
To better understand e-commerce integrations, it's crucial to consider their directionality, distinguishing between inbound and outbound integrations [29] [30].

\begin{itemize}
  \item \textit{Inbound Integrations:} SaaS-based e-commerce platforms [31] usually offer an integration layer based on Application Programming Interfaces (APIs). These APIs, typically RESTful web services [32][33], are exposed for third-party consumption. They are provided "out of the box," ready to be used by external requesters or legacy systems such as ERPs.
   \item \textit{Outbound Integrations:} These emanate from e-commerce platforms to an external entity and are usually executed through webhooks. These triggers send detailed order information to third-party entities such as the ERP system.
\end{itemize}

\subsubsection{The Role of AI in ECC and Third-Party Integrations}
As e-commerce evolves, secure and efficient third-party integrations are essential. ECC maintains data security and integrity across these integrations [34]. AI techniques optimize ECC parameters, boosting transaction speed, data security, and overall user experience, offering a competitive edge in e-commerce operations.

\subsection{Pre-Quantum Developments in ECC Optimization}
The advent of quantum computing ushers in a new era with its potential to solve complex problems more efficiently than classical computers [35]. However, the implications of this quantum leap for Elliptic Curve Cryptography (ECC) and its parameter optimization using AI algorithms remain largely speculative, as quantum computing is yet to become mainstream. The pre-quantum era, thus, serves as the current framework within which ECC optimization techniques are developed and implemented, focusing on the capabilities of classical computing.

\subsection{Limitations of Similar Research}
Although previous studies have made significant contributions to the field of ECC cryptanalysis and security, they primarily focus on specific techniques like Pollard's Rho, DNA-based cryptography, PSO/Cuckoo Search for key generation, and power optimization for mobile devices. These studies do not explore other AI techniques for optimizing ECC parameters, particularly in the context of e-commerce integrations. Furthermore, there is a lack of consideration for the practical applications of these optimizations in real-world scenarios. The limitations of these studies are summarized in Table \ref{table:Limitations_Similar_Research}.

\begin{table}[h]
    \renewcommand{\arraystretch}{1.2}
    \caption{Limitations of Similar Research}
    \label{table:Limitations_Similar_Research}
    \centering
    \begin{tabular}{|>{\centering\arraybackslash}p{1cm}|>{\centering\arraybackslash}p{6.5cm}|}
        \hline
        \textbf{Study} & \textbf{Limitations} \\
        \hline
        {[12]} & Focuses on cryptanalysis rather than optimization of ECC parameters for practical applications. Does not explore other AI techniques or their integration with e-commerce systems. \\
        \hline
        {[13]} & Concentrates on multi-cloud security using DNA and HECC techniques but does not explore other AI techniques like GA or PSO for ECC optimization. Lacks practical implementation details for e-commerce integrations. \\
        \hline
        {[16]} & Focuses on mobile devices and optimizing power consumption using PSO and Simplified Swarm Optimization. Does not provide a comprehensive comparison with other AI techniques like GA for ECC optimization. The study's focus on mobile device constraints limits its applicability to broader e-commerce integrations. \\
        \hline
        {[17]} & Explores PSO and Cuckoo Search Algorithm for ECC key selection but does not provide a comprehensive comparison with other AI techniques like GA. Focuses more on key generation rather than overall ECC parameter optimization in e-commerce contexts. \\
        \hline
    \end{tabular}
\end{table}

\section{Materials and Methods}
This section highlights our research methodology, focusing on the ECC parameters to optimize and the criteria for the evaluation of AI techniques.

\subsection{ECC Optimization Parameters}
Elliptic Curve Cryptography parameters play distinctive roles, and they can be carefully tuned to improve ECC without sacrificing security. The parameters that will be analyzed for this study are as follows:

\subsubsection{Choice of Elliptic Curve}
The curve's equation \(E: y^2 = x^3 + ax + b\) and specific constants \(a\) and \(b\) (curve coefficients) determine the system's efficiency and security.

\subsubsection{Field Size}
Represented by a prime number \((p)\), the field size affects security and computational load. Larger fields enhance security but need careful balancing with efficiency.

\subsubsection{Generator point \(G\)}
The method used for representing points \((x, y)\) on the curve affects computation speed.

\subsubsection{Scalar Multiplication}
Techniques like the Montgomery ladder [36] or sliding window method [37]  enhance ECC operations. The operation \(Q = kP\), where \(P\) is a point on the curve and \(k\) is scalar, can be optimized for efficiency.

\subsubsection{Group Order \(n\)}
This represents the number of points on the elliptic curve and plays a vital role in the security of the ECC system.

\subsubsection{Cofactor \(h\)}
The ratio between the number of points on the curve and the group order \(n\). It's essential in defining the subgroup that is used for cryptographic purposes.

The parameters mentioned above represent only a fraction of the many that can be considered [1] – [11]. Other aspects, such as Hash Function, Pairing Function, Random Number Generation, protocol parameters, use of special curves, batch operations, endomorphism \(( \phi: E \rightarrow E )\), parallelism, efficient arithmetic libraries, hardware acceleration, and more, will not be discussed to maintain the focus of the study.

\subsection{ECC Optimization Criteria}
The efficiency and effectiveness, collectively referred to as the efficacy, of the selected AI algorithms in optimizing the ECC parameters, are assessed based on multiple criteria acknowledged as vital evaluation measures by the broader research community. These criteria encompass various aspects that together represent the complete performance of ECC. Below, we outline these criteria:

\subsubsection{Evaluation of the AI Algorithms (efficiency)}
\subsubsubsection{Performance}
Speed, convergence rate, computational time.
\subsubsubsection{Flexibility}
Ability to adapt to different problems or changes in the landscape.
\subsubsubsection{Robustness}
Sensitivity to initial conditions, parameter settings, and noise.
\subsubsubsection{Scalability}
Ability to handle increasing complexity or problem size.
\subsubsubsection{Comparability}
Fairness and alignment in comparing the two algorithms.
\subsubsection{Evaluation of the ECC Parameters Generated (effectiveness)}
\subsubsubsection{Security}
Resistance against attacks, adherence to cryptographic best practices.
\subsubsubsection{Optimality}
How close the parameters are to the theoretical best solution.
\subsubsubsection{Generalization}
Effectiveness across different curve configurations and real-world scenarios.
\subsubsubsection{Validity}
Compliance with mathematical and cryptographic requirements, such as avoiding singular or anomalous curves.
\subsubsubsection{Practicality}
Consideration of real-world applications, computational performance, and compatibility with existing systems.

The two aspects of efficiency and effectiveness, are interconnected in efficacy but evaluate different dimensions of the problem. Efficiency focuses on the algorithms themselves and how they perform as optimization techniques [38] - [43], while effectiveness concentrates on the quality and characteristics of the ECC parameters they produce [1] - [11].

\section{Simulation Environment Design and Implementation}

\begin{figure}[h]
    \centering
    \includegraphics[width=1\linewidth]{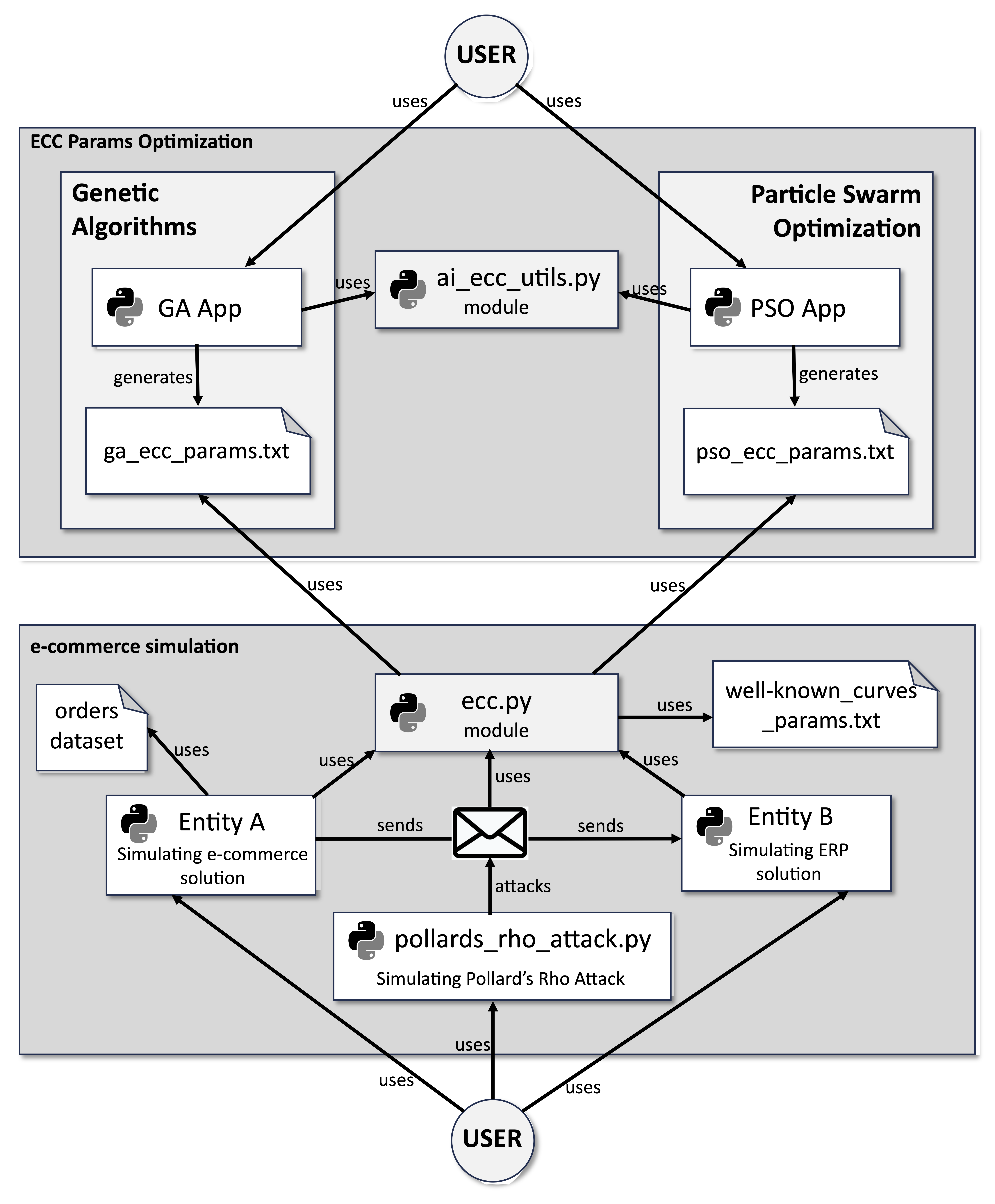}
    \caption{A high-level diagram of the environment's architecture}
    \label{fig:env_architecture}
\end{figure}

The implementation of our simulation consists of an environment of applications and software modules (hereafter referred to as components), built using the Python programming language. These are divided into two main groups: “ECC Params Optimization” and “e-commerce Simulation”. The architecture of this environment is illustrated in Figure \ref{fig:env_architecture}.

The libraries used in the project include, but are not limited to, `numpy`, `pandas`, `matplotlib`, `deap`, `gmpy2`, `requests`, and `tinyec`. These libraries, along with the detailed source code, can be found at: \href{https://github.com/cftellezc/GA_PSO_ECC_parameter_Optimization}{github.com/cftellezc/GA\_PSO\_ECC\_parameter\_Optimization}

The components of our simulation environment are described in more detail below.

\subsection{ECC params optimization group}

\subsubsection{Genetic Algorithm}
The GA.py script or GA App, employs the DEAP (Distributed Evolutionary Algorithms in Python) library to implement a genetic algorithm for ECC parameter optimization. It initiates a population of individuals, where each individual is a list representing potential elliptic curve parameters in ECC. This parameters are the constants \(a\) and \(b\), the prime number \(p\), the generator point \(G\) representing points \((x, y)\), the group order \(n\) and the cofactor \(h\). Through iterative genetic operations like selection, crossover, and mutation, new generations of individuals are produced.

The script uses a custom mutation function that receives the individual, an independent probability \textit{indpb} (the chance of each attribute to be mutated), and the mutation rate, all of this to mutate the individuals generating prime numbers or perturbing parameters using a Gaussian distribution as follows (Algorithm \ref{alg:customMutation}):

\begin{algorithm}
\caption{Custom Mutation Function}
\label{alg:customMutation}
\begin{algorithmic}[1]
\Function{customMutation}{individual, indpb, mutation\_rate}
    \State degree\_of\_mutation $\gets$ 5
    \If{mutation\_rate $>$ 0.5}
        \State degree\_of\_mutation $\gets$ 10
    \Else
        \State degree\_of\_mutation $\gets$ 2
    \EndIf
    \If{random value between 0 and 1 $<$ mutation\_rate}
        \For{i = 0 to length of individual $-$ 1}
            \If{random value between 0 and 1 $<$ indpb}
                \If{i = 2}
                    \State individual[i] $\gets$ generate prime number for p
                \ElsIf{individual[i] is a tuple}
                    \State individual[i] $\gets$ find generator point using individual[0], individual[1], individual[2]
                \Else
                    \State individual[i] $\gets$ individual[i] + round(value from Gaussian distribution with mean 0 and standard deviation degree\_of\_mutation)
                \EndIf
            \EndIf
        \EndFor
    \EndIf
    \State \Return individual,
\EndFunction
\end{algorithmic}
\end{algorithm}

The script defines several global constants that control the behavior of the genetic algorithm, as summarized in Table \ref{table:GA_Constants}:

\begin{table}[!htb]
    \renewcommand{\arraystretch}{1.3}
    \caption{Genetic Algorithm Constants}
    \label{table:GA_Constants}
    \centering
    \begin{tabular}{|c|c|c|}
        \hline
        \textbf{Constant} & \textbf{Value} & \textbf{Description} \\
        \hline
        POP\_SIZE & 500 & Population size \\
        \hline
        CXPB & 0.5 & Crossover probability \\
        \hline
        MUTPB & 0.2 & Mutation probability \\
        \hline
        NGEN & 40 & Number of generations \\
        \hline
        MULTIPARENT\_CXPB & 0.1 & Multi-parent crossover probability \\
        \hline
        ELITISM\_RATE & 0.1 & Elitism rate \\
        \hline
    \end{tabular}
\end{table}

\newpage
They were set up and tuned for better performance using some techniques like grid search. They can be adjusted to tune the performance of the algorithm.

Executed as the main module, the script's primary function initializes the population, assesses their fitness, and enters a loop for generating and evaluating new individuals over a specified number of generations (\textit{NGEN}) as follows (Algorithm \ref{alg:mainGA}):

\begin{algorithm}
\caption{Main Genetic Algorithm}
\label{alg:mainGA}
\begin{algorithmic}[1]
\State \textbf{Initialize:} $pop \leftarrow \text{toolbox.population()}$

\ForAll{individual in $pop$}
    \State Evaluate fitness and assign to individual
\EndFor

\State \textbf{Initialize:} $elitism\_number \leftarrow \text{round}(len(pop) \times ELITISM\_RATE)$
\State $mutation\_rate \leftarrow MUTPB$

\For{$g=0$ to $NGEN-1$}
    \State \text{Log: Starting Generation: $g+1$}
    \State $elites \leftarrow \text{select top individuals from pop}$
    \State $offspring \leftarrow \text{select next generation from pop}$
    \State \textbf{Clone:} offspring
    
    \For{$i=0$ to $len(offspring) - 3$ \textbf{ step } 3}
        \State Apply three-point crossover if random() $< MULTIPARENT\_CXPB$
    \EndFor
    
    \ForAll{pair $child1, child2$ in offspring}
        \State Apply crossover to $child1, child2$ if random() $< CXPB$
    \EndFor
    
    \ForAll{mutant in offspring}
        \State \textbf{mutate} mutant with rate $mutation\_rate$
    \EndFor
    
    \State Evaluate and set fitness of invalid individuals
    \State $offspring \leftarrow offspring + elites$
    \State $pop \leftarrow offspring$
\EndFor
\end{algorithmic}
\end{algorithm}

Summarizing, we initialize the mutation degree utilizing the standard deviation of a Gaussian distribution, and adjusting it according to the \textit{mutation\_rate}. This procedure entails creating random floats to determine the probability of mutation, which is influenced by both the mutation and \textit{indpb} rates.

Individual parameters are subject to potential mutation, adhering to unique criteria for distinct cases: the third parameter ($p$) entails deriving a fresh prime number utilizing the \textit{BITS\_PRIME\_SIZE} constant — currently set to 256 bits — from the \textit{ai\_ecc\_utils.py} module. Tuple parameters representing generator points ($G$) necessitate fabricating a new point using the prevailing values of $a$, $b$, and $p$.

The remaining parameters ($a$ and $b$) undergo modifications through Gaussian perturbations, with an imperative to retain integer attributes facilitated by the round function. This mechanism ensures compatibility with DEAP, expecting mutated individuals to be returned as single-item tuples.

The script uses tournament selection and two-point crossover from the DEAP library. It also implements elitism, ensuring that the best individuals from each generation are carried over to the next. 

The script evaluates the fitness of the individuals using a function from the \textit{ai\_ecc\_utils.py} module (which will be explained later), which calculates the fitness based on the ECC parameters represented by the individual. The script logs the progress of the genetic algorithm, including the statistics of each generation and the best individual from the final generation. 

In the context of \textit{ECC}, the individuals in the population represent different elliptic curves, and the fitness function assesses how well they meet the desired criteria. The genetic algorithm identifies the best-fitting elliptic curve and generates a file named \textit{ga\_ecc\_params.txt}, containing the optimal parameters for \textit{ECC} optimization.

\subsubsection{Particle Swarm Optimization}
The PSO.py module or PSO App, uses a Particle Swarm Optimization (PSO) algorithm to fine-tune ECC parameters. It initializes particles, where each particle is a list representing potential elliptic curve parameters in ECC, updates their velocities and positions, evaluates fitness, and identifies the best ECC parameter set through the optimal particle. The ECC parameters are the same as those assessed in GA: \((a, b, p, G, n, h)\).

The \textit{update\_velocity} function calculates the new velocity of a particle. It balances global and local exploration using a dynamic inertia weight that linearly decreases from $0.9$ to $0.4$ over iterations. Two components contribute to the velocity: a cognitive component based on the particle's best-known position $C1$, and a social component based on the swarm's best-known position $C2$. Special handling is done for the generator point of the elliptic curve, as shown in Algorithm \ref{alg:updateVelocity}.

\begin{algorithm}
\caption{Update Velocity of a Particle}
\label{alg:updateVelocity}
\begin{algorithmic}[1]
\small
\Procedure{UpdateVel}{part, vel, best\_part, glob\_best\_part, iter, max\_iter}
    \State $new\_vel \gets \text{{[]}}$
    \State $w_{\text{{max}}} \gets 0.9$
    \State $w_{\text{{min}}} \gets 0.4$
    \State $w \gets w_{\text{{max}}} - (w_{\text{{max}}} - w_{\text{{min}}}) \cdot \frac{{\text{{iter}}}}{{\text{{max\_iter}}}}$
    \For{$i = 0$ \textbf{to} $\text{{len(part)}} - 1$}
        \State $v \gets \text{{vel}}[i]$
        \State $r1, r2 \gets \text{{random(0, 1)}}$
        \If{$i = 3$}
            \State $cog \gets \text{{calc\_cog}}(best\_part[i], part[i], r1)$
            \State $soc \gets \text{{calc\_soc}}(glob\_best\_part[i], part[i], r2)$
            \If{$\text{{is\_tuple}}(v)$}
                \State $new\_v \gets \text{{calc\_new\_v\_tuple}}(v, w, cog, soc)$
            \Else
                \State $new\_v \gets \text{{calc\_new\_v}}(cog, soc)$
            \EndIf
        \Else
            \State $cog \gets C1 \cdot r1 \cdot (best\_part[i] - part[i])$
            \State $soc \gets C2 \cdot r2 \cdot (glob\_best\_part[i] - part[i])$
            \State $new\_v \gets w \cdot v + cog + soc$
            \If{$i \neq 3$}
                \State $new\_v \gets \text{{abs(new\_v)}}$
            \EndIf
        \EndIf
        \State \textbf{Append} $new\_v$ \textbf{to} $new\_vel$
    \EndFor
    \State \textbf{return} $new\_vel$
\EndProcedure
\end{algorithmic}
\end{algorithm}

\newpage
The \textit{update\_position} function calculates the new position of a particle based on its velocity. It ensures that coordinates remain positive and integers, and specific attention is given to update the generator point. It also makes use of the external utility functions from \textit{ai\_ecc\_utils.py} to update the prime number and generator point, as shown in Algorithm \ref{alg:updatePosition}.

\begin{algorithm}
\caption{Update Position of a Particle}
\label{alg:updatePosition}
\begin{algorithmic}[1]
\Procedure{UpdatePosition}{particle, velocity}
    \State new\_particle $\gets$ empty list
    \For{$i = 0$ to $\text{len(particle)} - 1$}
        \State $p \gets \text{particle}[i]$
        \State $v \gets \text{velocity}[i]$
        \If{$i = 3$}
            \State new\_p $\gets$ tuple($\left| \text{int}\left(\text{round}(p_i + v_i)\right)\right|$ for $p_i, v_i$ in zip(p[:2], v[:2]))
        \Else
            \State new\_p $\gets \left|\text{int}\left(\text{round}(p + v)\right)\right|$
        \EndIf
        \State Append new\_p to new\_particle
    \EndFor
    \State new\_particle[2] $\gets$ ai\_ecc\_utils.get\_prime\_for\_p()
    \State a, b, p $\gets$ new\_particle[:3]
    \State new\_particle[3] $\gets$ ai\_ecc\_utils.find\_generator\_point(a, b, p)
    \State \Return new\_particle
\EndProcedure
\end{algorithmic}
\end{algorithm}

The script specifies several global constants that regulate how the Particle Swarm Optimization algorithm behaves, as summarized in Table \ref{table:PSO_Constants}:

\begin{table}[!htb]
    \renewcommand{\arraystretch}{1.3}
    \caption{Particle Swarm Optimization Constants}
    \label{table:PSO_Constants}
    \centering
    \begin{tabular}{|p{0.3\columnwidth}|c|p{0.4\columnwidth}|}
        \hline
        \centering\textbf{Constant} & \centering\textbf{Value} & \centering\textbf{Description} \tabularnewline
        \hline
        SWARM\_SIZE & 500 & Number of particles in the swarm\tabularnewline
        \hline
        MAX\_ITERATIONS & 40 & Maximum number of iterations\tabularnewline
        \hline
        C1 & 1.0 & Cognitive parameter (influence of particle's best-known position)\tabularnewline
        \hline
        C2 & 2.5 & Social parameter (influence of swarm's best-known position)\tabularnewline
        \hline
        MAX\_ITERATIONS\_ WITHOUT\_ IMPROVEMENT & 20 & Used for an early stopping feature\tabularnewline
        \hline
    \end{tabular}
\end{table}

The implementation includes a parameter grid for tuning the PSO constants.

The \textit{main function} initializes the swarm of particles, their velocities, and their best-known positions. Global best-known positions are also identified. The \textit{main loop} iterates through the swarm, updating velocities and positions using the previously defined functions. Fitness is evaluated for each particle using the same fitness function from the \textit{ai\_ecc\_utils.py} module that is used by \textit{GA.py} (which will be explained later), and best-known positions are updated as necessary. If there is no improvement in global best fitness for \(20\) iterations, the algorithm stops early, and at the end, statistics regarding fitness values are calculated and printed. The best particle is selected, and its details are printed and written to the \textit{pso\_ecc\_params.txt} file.

\subsubsection{ga\_ecc\_params.txt}
This file contains the best parameters found by the Genetic Algorithm (GA) for ECC parameter optimization.

\subsubsection{pso\_ecc\_params.txt}
Represents the file with the best ECC parameters found by the Particle Swarm Optimization (PSO) technique.

\subsubsection{ai\_ecc\_utils.py}
It is a utility module that aids AI algorithms like GA and PSO in the process of ECC parameter optimization. The module's primary purpose is to facilitate the creation of elliptic curves and their associated parameters, as shown in Algorithm \ref{alg:generateCurve}.

\begin{algorithm}
\caption{Elliptic Curve Parameter Generation}
\label{alg:generateCurve}
\begin{algorithmic}[1]
\Procedure{generate\_curve}{}
    \State $\text{signal.signal}(\text{signal.SIGALRM}, \text{handler})$
    \While{True}
        \State $p \gets \text{get\_prime\_for\_p()}$
        \While{True}
            \State $\text{logging.info}("a, b \text{ generation}")$
            \State $a \gets \text{random.randint}(0, p - 1)$
            \State $b \gets \text{random.randint}(0, p - 1)$
            \If{$(4 \cdot a^3 + 27 \cdot b^2) \mod p \neq 0$ and $\text{not is\_singular}(a, b, p)$}
                \State break
            \EndIf
        \EndWhile
        \State \textbf{try:}
            \State $\text{signal.alarm}(\text{TIMEOUT\_SECONDS})$
            \State $G \gets \text{find\_generator\_point}(a, b, p)$
            \State $\text{logging.info}("G: ", G)$
            \State $\text{signal.alarm}(0)$
            \State break
        \State \textbf{except} $\text{NoGeneratorPointException}, \text{TimeoutError}:$
            \State continue
    \EndWhile
    \State $n \gets p - 1$
    \State $h \gets 1$
    \State \Return $(a, b, p, G, n, h)$
\EndProcedure

\Procedure{get\_prime\_for\_p}{}
    \State \Return $\text{getPrime}(\text{BITS\_PRIME\_SIZE})$
\EndProcedure

\Procedure{is\_singular}{a, b, p}
    \State $\text{discriminant} \gets (4 \cdot a^3 + 27 \cdot b^2) \mod p$
    \State \Return $\text{discriminant} == 0$
\EndProcedure

\Procedure{find\_generator\_point}{a, b, p}
    \For{$x$ in $0$ to $p - 1$}
        \State $\text{rhs} \gets (x^3 + a \cdot x + b) \mod p$
        \If{$\text{legendre\_symbol}(\text{rhs}, p) == 1$}
            \State $y \gets \text{tonelli\_shanks}(\text{rhs}, p)$
            \State \Return $(x, y)$
        \EndIf
    \EndFor
    \State \textbf{raise} $\text{NoGeneratorPointException}$
\EndProcedure

\end{algorithmic}
\end{algorithm}

It includes functions for generating prime numbers and finding generator points on the elliptic curve using mathematical functions, such as the Legendre symbol and Tonelli-Shanks algorithm [44]. The Legendre symbol determines whether a number is a quadratic residue modulo a prime, essential for finding valid points on the elliptic curve. The Tonelli-Shanks algorithm finds the square root of a number modulo a prime, crucial for computing the y-coordinates of the points on the curve. These methods ensure the generated points are valid and lie on the elliptic curve, as shown in Algorithm \ref{alg:legendre_tonelli}.

\begin{algorithm}
\caption{Legendre Symbol and Tonelli-Shanks Algorithm}
\label{alg:legendre_tonelli}
\begin{algorithmic}[1]
\Procedure{legendre\_symbol}{a, p}
    \State $ls \gets \text{pow}(a, (p - 1) \div 2, p)$
    \State \Return $-1$ if $ls == p - 1$ else $ls$
\EndProcedure

\Procedure{tonelli\_shanks}{n, p}
    \State $\text{assert } \text{legendre\_symbol}(n, p) == 1, \text{"n is not a quadratic residue modulo p"}$
    \State $q \gets p - 1$
    \State $s \gets 0$
    \While{$q \mod 2 == 0$}
        \State $q \div= 2$
        \State $s \gets s + 1$
    \EndWhile
    \If{$s == 1$}
        \State \Return $\text{pow}(n, (p + 1) \div 4, p)$
    \EndIf
    \For{$z$ from $2$ to $p - 1$}
        \If{$\text{legendre\_symbol}(z, p) == -1$}
            \State break
        \EndIf
    \EndFor
    \State $m \gets s$
    \State $c \gets \text{pow}(z, q, p)$
    \State $t \gets \text{pow}(n, q, p)$
    \State $r \gets \text{pow}(n, (q + 1) \div 2, p)$
    \While{$t \neq 1$}
        \State $i \gets 0$
        \State $t_i \gets t$
        \While{$t_i \neq 1$}
            \State $t_i \gets \text{pow}(t_i, 2, p)$
            \State $i \gets i + 1$
        \EndWhile
        \State $b \gets \text{pow}(c, 2^{m - i - 1}, p)$
        \State $r \gets r \cdot b \mod p$
        \State $t \gets t \cdot b \cdot b \mod p$
        \State $c \gets b \cdot b \mod p$
        \State $m \gets i$
    \EndWhile
    \State \Return $r$
\EndProcedure
\end{algorithmic}
\end{algorithm}

The \textit{ai\_ecc\_utils.py} module validates ECC parameters, checking cofactor, prime \(p\), point validity, handling generator point exceptions, matching order and cofactor, and confirming non-singular, anomalous, or supersingular characteristics [45], as shown in Algorithm \ref{alg:validateCurve}. These validations ensure the integrity and security of the ECC parameters used in the cryptographic system.

\begin{algorithm}
\caption{Validation and Properties of Elliptic Curve}
\label{alg:validateCurve}
\begin{algorithmic}[1]

\Procedure{validate\_curve}{$a, b, p, G, n, h$}
    \If{$h < 1$}
        \State \textbf{log} "The cofactor h is less than 1, which makes it invalid."
        \State \Return False
    \EndIf
    \If{$p == 0$}
        \State \textbf{log} "The prime p can't be zero."
        \State \Return False
    \EndIf
    \If{len($G$) == 2}
        \State $x, y \gets G$
        \If{$(y \cdot y - x \cdot x \cdot x - a \cdot x - b) \mod p \neq 0$}
            \State \textbf{log} "The point G is not on the curve!"
            \State \Return False
        \EndIf
        \State $field \gets$ SubGroup with $(p, G, n, h)$
        \If{No generator point in $field$}
            \State \textbf{log} "No generator point found!"
            \State \Return False
        \EndIf
        \State $curve \gets$ Curve with $(a, b, field,$ \text{"random\_curve"})
    \Else
        \State \textbf{log} "Invalid generator point provided. Skipping curve creation."
        \State \Return False
    \EndIf
    \State $order \gets n$
    \If{$h \neq field.h$}
        \State \textbf{log} "The cofactor does not match the expected cofactor!"
        \State \Return False
    \EndIf
    \If{\Call{is\_singular}{$a, b, p$}}
        \State \textbf{log} "The curve is singular!"
        \State \Return False
    \EndIf
    \If{\Call{is\_anomalous}{$p, order$}}
        \State \textbf{log} "The curve is anomalous!"
        \State \Return False
    \EndIf
    \If{\Call{is\_supersingular}{$p, order$}}
        \State \textbf{log} "The curve is supersingular!"
        \State \Return False
    \EndIf
    \State \Return True
\EndProcedure

\Procedure{is\_singular}{$a, b, p$}
    \State $discriminant \gets (4 \cdot a^3 + 27 \cdot b^2) \mod p$
    \State \Return $discriminant == 0$
\EndProcedure

\Procedure{is\_anomalous}{$p, n$}
    \State \Return $p == n$
\EndProcedure

\Procedure{is\_supersingular}{$p, n$}
    \If{$p \in [2, 3]$ or not isprime($p$)}
        \State \Return False
    \EndIf
    \State \Return $(p + 1 - n) \mod p == 0$
\EndProcedure

\end{algorithmic}
\end{algorithm}

The module also implements Pollard's rho attack [45] to evaluate the security of the generated ECC parameters. This attack is a well-known method for finding discrete logarithms in elliptic curves, making it an essential tool for assessing the resilience of the cryptographic system against specific types of attacks. It employs functions to add two points on an elliptic curve, apply the "double and add" method for point multiplication, and check if a point is "distinguished" by having \(t\) trailing zeros in its \textit{x-coordinate}, as shown in Algorithm \ref{alg:pollardRho}.

\begin{algorithm}
\caption{Pollard's Rho Attack on an Elliptic Curve}
\label{alg:pollardRho}
\begin{algorithmic}[1]
\Function{p\_rho\_attack}{$G, a, b, p, \text{{order}}, t, \text{{max\_iter}}$}
    \State $Q_a, Q_b \gets G, G$
    \State $a, b \gets 0, 0$
    \State $\text{{power\_of\_two}} \gets 1$
    \State $\text{{iterations}} \gets 0$
    \While{$\text{{iterations}} < \text{{max\_iterations}}$}
        \For{$\_ \text{{ in range}}(\text{{power\_of\_two}})$}
            \State $i \gets Q_a[0] \mod 3$
            \If{$i = 0$}
                \State $Q_a \gets \text{{add\_points}}(Q_a, G, a, p)$
                \State $a \gets (a + 1) \mod \text{{order}}$
            \ElsIf{$i = 1$}
                \State $Q_a \gets \text{{double\_and\_add}}(2, Q_a, a, p)$
                \State $a \gets (2 \cdot a) \mod \text{{order}}$
            \Else
                \State $Q_a \gets \text{{double\_and\_add}}(2, Q_a, a, p)$
                \State $a \gets (2 \cdot a) \mod \text{{order}}$
                \State $Q_a \gets \text{{add\_points}}(Q_a, G, a, p)$
                \State $a \gets (a + 1) \mod \text{{order}}$
            \EndIf
            \If{$\text{{is\_distinguished}}(Q_a, t)$}
                \State \Return $a, Q_a$
            \EndIf
        \EndFor
        \For{$\_ \text{{ in range}}(2)$}
            \State \text{{Repeat the same steps for }}$Q_b$
            \State \text{{ , but twice per iteration}}
        \EndFor
        \State $\text{{iterations}} \gets \text{{iterations}} + 1$
        \If{$Q_a = Q_b$}
            \State $\text{{power\_of\_two}} \gets \text{{power\_of\_two}} \times 2$
            \State $Q_b \gets Q_a$
            \State $b \gets a$
        \EndIf
    \EndWhile
    \State $\text{{logging.info}}(''\text{{No collision found within the}})$
    \State $\text{{specified maximum number of iterations.}}'')$
    \State \Return \text{{None}}
\EndFunction
\end{algorithmic}
\end{algorithm}

Lastly, the \textit{ai\_ecc\_utils.py} module calculates the fitness function, incorporating all the elliptic curve validations. It evaluates the fitness of a candidate, whether an “individual” in the GA population or a “particle” in the PSO swarm, as shown in Algorithm \ref{alg:evaluateFitness}.

\begin{algorithm}
\caption{Function to evaluate the fitness of a candidate in GA or PSO}
\label{alg:evaluateFitness}
\begin{algorithmic}[1]
\footnotesize
\Function{evaluate}{candidate}
    \State Extract $a, b, p, G, n, h$ from candidate
    \If{not \Call{validate\_curve}{$a, b, p, G, n, h$}}
        \State \Return 0
    \EndIf
    \State $expected\_order \gets p + 1 - 2 \cdot \sqrt{p}$
    \State $upper\_bound \gets expected\_order + 2 \cdot \sqrt{p}$
    \State $hasse\_score \gets \max\left(0, \frac{upper\_bound - \left|n - expected\_order\right|}{upper\_bound - lower\_bound}\right)$
    \State $start\_time \gets \text{current time}$
    \State $rho\_attack\_result \gets \Call{p\_rho\_attack}{G, a, b, p, expected\_order}$
    \State $execution\_time \gets \text{current time} - start\_time$
    \State $max\_time \gets 10.0, min\_time \gets 0.1$
    \State $execution\_score \gets \max\left(0, \min\left(1, \frac{execution\_time - min\_time}{max\_time - min\_time}\right)\right)$
    \State $attack\_resistance\_score \gets 1$ if $rho\_attack\_result$ is None else 0
    \State $fitness \gets 0.4 \cdot \log(n) + 0.2 \cdot hasse\_score \cdot \log(n) + 0.2 \cdot execution\_score + 0.2 \cdot attack\_resistance\_score$
    \State \Return fitness
\EndFunction
\end{algorithmic}
\end{algorithm}

The fitness function extracts the elliptic curve parameters \((a, b, p, G, n, h)\). \textit{Curve Validation} checks if the parameters form a valid curve, returning a fitness of \(0\) if not. The expected order of the curve is calculated, checks Hasse's theorem bounds [44] and the Hasse score is computed to evaluate how close the actual order is to the expected. \textit{Pollard's Rho Attack} is attempted, with a longer execution time indicating higher resistance. An \textit{Attack Resistance Score} is assigned. The \textit{Final Fitness Calculation} includes \(40\%\) weight to the natural logarithm of the curve's order, \(20\%\) to the Hasse score (weighted by the logarithm of the order), \(20\%\) to the execution time score of the attack, and \(20\%\) to the resistance score. The cumulative fitness score is returned, reflecting the candidate's elliptic curve suitability.

These are several global constants that are defined for \textit{ai\_ecc\_utils.py} module, as summarized in Table \ref{table:ai_ecc_utils_constants}:

\begin{table}[!htb]
    \renewcommand{\arraystretch}{1.3}
    \caption{Ai\_ecc\_utils.py module Constants}
    \label{table:ai_ecc_utils_constants}
    \centering
    \begin{tabular}{|p{0.3\columnwidth}|c|p{0.4\columnwidth}|}
        \hline
        \centering\textbf{Constant} & \centering\textbf{Value} & \centering\textbf{Description} \tabularnewline
        \hline
        BITS\_PRIME\_SIZE & 256 & Size of the prime in bits. Generates a n-bit prime number\tabularnewline
        \hline
        POLLARDS\_RHO\_TRIALS & 20 & Number of trials for the Pollard's Rho function\tabularnewline
        \hline
        POLLARDS\_RHO\_ MAX\_ITER & 10**2 & Maximum iterations for each trial in the Pollard's Rho function\tabularnewline
        \hline
     \end{tabular}
\end{table}

This module is designed to be used in conjunction with other modules that implement bio-inspired algorithms, such as PSO and GA, and given how it was designed, other AI algorithms that are capable of optimizing elliptical curves may use it in the future.

\subsection{E-commerce Simulation group}

\subsubsection{well-known\_curves\_params.txt}
Represents parameters for standard elliptic curves used in cryptography, such as \textit{secp256k1} and \textit{brainpoolP256r1} [46].

\subsubsection{ecc.py}
This utility module includes classes and functions for elliptic curve cryptography. A key function reads ECC parameters from files like \textit{ga\_ecc\_params.txt}, \textit{pso\_ecc\_params.txt}, \textit{secp256k1.txt}, or \textit{brainpoolP256r1.txt} (the latter two as \textit{well-known\_curves\_params.txt)}, creating a structured set of parameters to be used in the e-commerce simulation, as shown in Algorithm \ref{alg:initializeParams}.

\begin{algorithm}
\caption{Initialization of ECC Parameters}
\label{alg:initializeParams}
\begin{algorithmic}[1]
\Function{initialize\_params}{option}
    \State Select $filename$ based on $option$:
    \If{option = "1"} 
        \State $filename \gets "ga\_ecc\_params.txt"$ 
    \ElsIf{option = "2"} 
        \State $filename \gets "pso\_ecc\_params.txt"$ 
    \ElsIf{option = "3"} 
        \State $filename \gets "secp256k1.txt"$ 
    \ElsIf{option = "4"} 
        \State $filename \gets "brainpoolP256r1.txt"$ 
    \Else 
        \State $filename \gets "secp256k1.txt"$ \Comment{default}
    \EndIf
    \State Open $filename$ for reading as $f$
    \State Initialize $params\_dict$ as an empty dictionary
    \For{each line in $f$}
        \State Split line into $key$, $value$ and store in $params\_dict$
        \State Convert $value$ to integer
    \EndFor
    \State Extract parameters $p, a, b, G_x, G_y, n, h$ from $params\_dict$
    \State $G \gets \text{ECPoint}(G_x, G_y)$
    \State \Return $\text{ECCParameters}(p, a, b, G, n, h)$
\EndFunction
\end{algorithmic}
\end{algorithm}

The \textit{ecc.py} module has support functions like \textit{ec\_addition} for adding points on an elliptic curve, and \textit{ec\_scalar\_multiplication} for multiplying a point by a scalar using the double-and-add method, as shown in Algorithm \ref{alg:pointAddition}.

\begin{algorithm}
\caption{Point Addition and Scalar Multiplication}
\label{alg:pointAddition}
\begin{algorithmic}[1]
\Function{ec\_addition}{$P, Q, p$}
    \If{$P$ is None or inf} \Return $Q$ \EndIf
    \If{$Q$ is None or inf} \Return $P$ \EndIf
    \If{$P.x = Q.x$}
        \If{$P.y = -Q.y \mod p$} 
            \Return ECPoint(None, None) \Comment{Infinity}
        \EndIf
        \State $m \gets (3P.x^2 + a) \cdot (2P.y)^{-1} \mod p$
    \Else
        \State $m \gets (Q.y - P.y) \cdot (Q.x - P.x)^{-1} \mod p$
    \EndIf
    \State $x \gets m^2 - P.x - Q.x \mod p$
    \State $y \gets m(P.x - x) - P.y \mod p$
    \Return ECPoint($x, y$)
\EndFunction

\Function{ec\_scalar\_mul.}{$P, s, p$}
    \State $r \gets$ ECPoint(None, None) 
    \State $c \gets P$
    \While{$s$}
        \If{$s \& 1$} 
            \State $r \gets$ is None ? $c :$ ec\_addition$(r, c, p)$ 
        \EndIf
        \State $c \gets$ ec\_addition$(c, c, p)$
        \State $s \mathrel{>>}= 1$
    \EndWhile
    \State Print $r.x, r.y$
    \Return $r$
\EndFunction
\end{algorithmic}
\end{algorithm}

These utility functions facilitate key generation by creating a random private key, an integer within the range \([1, n-1]\), and producing the corresponding public key, which is computed by multiplying the base point \(G\) by the private key, as shown in Algorithm \ref{alg:keyGeneration}.

\begin{algorithm}
\caption{Private and Public Key Generation}
\label{alg:keyGeneration}
\begin{algorithmic}[1]

\Function{generate\_private\_key}{$\text{params}$}
    \State \Return $\text{randbelow}(\text{params}.n - 1)$
\EndFunction

\Function{generate\_public\_key}{$\text{private\_key}, \text{params}$}
   \State $result \gets \text{ec\_scalar\_multiplication}(\text{params}.G,$ \\
  \hspace*{2.5em}$\text{private\_key},\text{params})$
  \State \Return $result$
\EndFunction

\end{algorithmic}
\end{algorithm}

Additionally, within the \textit{ecc.py} module, the functions \textit{ec\_addition} and \textit{ec\_scalar\_multiplication} are utilized to encrypt and decrypt messages, as shown in Algorithm \ref{alg:eccEncryption}.

\begin{algorithm}
\caption{ECC Encryption and Decryption}
\label{alg:eccEncryption}
\begin{algorithmic}[1]

\Function{encrypt\_message}{$\text{message}, \text{public\_key}, \text{params}$}
    \If{not $\text{is\_valid\_point}(\text{public\_key}, \text{params})$}
        \State \textbf{raise} ValueError("Public key is not a valid point on the elliptic curve")
    \EndIf
    \State $k \gets \text{generate\_private\_key}(\text{params})$
    \If{not $\text{is\_valid\_scalar}(k, \text{params})$}
        \State \textbf{raise} ValueError("Invalid scalar value")
    \EndIf
    \State $C1 \gets \text{ec\_scalar\_multiplication}(\text{params}.G, k, \text{params})$
    \State $C2 \gets \text{ec\_scalar\_multiplication}(\text{public\_key}, k, \text{params})$
    \State $\text{message\_bytes} \gets \text{message}.encode('utf-8')$
    \State $\text{encrypted\_message} \gets []$
    \For{byte in $\text{message\_bytes}$}
        \If{$C2.x$ is None}
            \State \textbf{raise} ValueError("Encryption failed: kQ resulted in the point at infinity")
        \EndIf
        \State $\text{encrypted\_byte} \gets \text{byte} \oplus (C2.x \& 0\text{xFF})$
        \State $\text{encrypted\_message.append}(\text{encrypted\_byte})$
    \EndFor
    \State \Return $C1, \text{encrypted\_message}$
\EndFunction
\Function{decrypt\_message}{$C1, \text{encrypted\_message},$}
   \hspace*{2.5em}$\text{private\_key},\text{params})$
    \State \Comment{Try block starts here}
    \State $C2 \gets \text{ec\_scalar\_multiplication}(C1, \text{private\_key}, \text{params})$
    \State \Comment{Catch block starts here}
    \State \textbf{print}(f"An error occurred during decryption: \{str(e)\}")
    \State \Return None
    \State \Comment{Catch block ends here}
    \State $\text{decrypted\_message\_bytes} \gets \text{bytearray}()$
    \For{$\text{encrypted\_byte}$ in $\text{encrypted\_message}$}
        \State $\text{byte} \gets \text{encrypted\_byte} \oplus (C2.x \& 0\text{xFF})$
        \State $\text{decrypted\_message\_bytes.append}(\text{byte})$
    \EndFor
    \State $\text{decrypted\_message} \gets \text{decrypted\_message\_bytes}.decode('utf-8')$
    \State \Return $\text{decrypted\_message}$
    \State \Comment{Try block ends here}
\EndFunction

\end{algorithmic}
\end{algorithm}

Lastly, \textit{ecc.py} uses the \textit{hmac} Python library to generate and verify a Hash-based Message Authentication Code (HMAC) for a given message and key, ensuring the integrity and authenticity of messages.

\subsubsection{Orders dataset}
This e-commerce dataset includes invoices generated by an authorized online retailer [47]. These have been pre-processed and curated as orders for practical simulation purposes, converting them into order data. This data feeds \textit{Entity A} and is then sent to \textit{Entity B} in the e-commerce simulation. 

\subsubsection{EntityA.py}
This component emulates an e-commerce solution, and communicates with a simulated ERP server (\textit{EntityB.py}) via Elliptic Curve Cryptography (ECC). It includes functionalities for server connections, ECC parameters, key management, order message generation, and transaction handling. The system reads data from a spreadsheet file (Orders dataset), encrypts messages with ECC and HMAC using the module \textit{ecc.py}, sends them to the ERP server, and employs Elliptic Curve Diffie-Hellman (ECDH) for key agreement, as shown in Algorithm \ref{alg:mainEntityA}.

\begin{algorithm}
\caption{Main Function Procedure for Entity A}
\label{alg:mainEntityA}
\begin{algorithmic}[1]
\Procedure{main}{}
    \State Initialize ServerConnection with server URL
    \State Initialize ECCParams with ServerConnection
    \State Request ECC parameters from server
    \State Initialize ECCKeys with ECC parameters
    \State Generate ECC private and public keys
    \State Initialize RetailMessage with MS Excel file path
    \State Initialize TransactionManager with necessary objects
    \State Run transactions until a predetermined end time
\EndProcedure
\end{algorithmic}
\end{algorithm}

\subsubsection{EntityB.py}
It is a simulated ERP server that interacts with the emulated e-commerce solution (\textit{EntityA.py}) using \textit{Flask} (Python web framework). It handles orders, key retrievals, initiates ECC and ECDH keys, and allows users to select the type of ECC parameters (such as GA, PSO, or well-known curves) to be used throughout the simulation. Functions for decrypting orders and verifying HMACs are included.

The ECC parameters employed in the simulation are selected by the user and loaded from the corresponding txt file. The \textit{ecc.py} module handles all cryptographic operations throughout the process.

\subsubsection{pollards\_rho\_attack.py}
To evaluate the ECC parameters in the simulation, a component is designed to attack the communication between \textit{EntityA.py} and \textit{EntityB.py}. This Python script executes \textit{Pollard's rho attack} [36][45] on the e-commerce simulation, employing the \textit{"tortoise and hare"} technique to implement the attack logic. By leveraging multiprocessing for parallelization and handling collisions to determine the private key, it interacts with the ERP server (\textit{EntityB.py}) to gather essential data such as the public key of the targeted entity. The script employs various methods like scalar multiplication and point addition on elliptic curve points to successfully carry out the attack, as shown in Algorithm \ref{alg:pollardsRhoAttack}.

It is important to note that within the \textit{ai\_ecc\_utils.py} component, there is a function that leverages the logic of Pollard's Rho attack to evaluate ECC parameters. While this function shares core mathematical principles with those in \textit{pollard\_rho\_attack.py}, their objectives differ significantly. Specifically, the function in the first code seeks to find collisions in points to assess the security of ECC and facilitate fitness calculation. The second code aims to find the private key. It carries additional logic to compute the private key from the collision, and employs the tortoise and hare approach, standard in Pollard's rho. Its goal is to attack the e-commerce scenario by finding the private key, a computationally challenging task.

\begin{algorithm}
\caption{Pollard's Rho Attack on Elliptic Curve Cryptography}
\label{alg:pollardsRhoAttack}
\begin{algorithmic}[1]
\Procedure{p\_rho}{\small{$\text{init\_value, G, public\_key, params, manager\_dict}$}}
    \State Print start message with init\_value
    \State tortoise $\gets$ ec\_scalar\_multiplication(G, init\_value, params)
    \State hare $\gets$ tortoise
    \State tortoise\_scalar $\gets$ hare\_scalar $\gets$ init\_value
    \For{i = 1 to $2 \times \text{params.n} + 1$}
        \State tortoise, tortoise\_scalar $\gets$ step(tortoise, tortoise\_scalar, G, public\_key, params)
        \State hare, hare\_scalar $\gets$ step(hare, hare\_scalar, G, public\_key, params)
        \State hare, hare\_scalar $\gets$ step(hare, hare\_scalar, G, public\_key, params)
        \If{tortoise = hare and tortoise $\neq$ None}
            \State scalar\_difference $\gets$ gmpy2.f\_mod((tortoise\_scalar - hare\_scalar), params.n)
            \If{scalar\_difference = 0}
                \State \textbf{continue}
            \EndIf
            \State scalar\_difference\_inverse $\gets$ gmpy2.invert(scalar\_difference, params.n)
            \State secret\_key $\gets$ gmpy2.f\_mod((scalar\_difference\_inverse * hare\_scalar), params.n)
            \If{$\neg$ manager\_dict['found\_flag']}
                \State manager\_dict['found\_flag'] $\gets$ True
                \State Print found secret key
            \EndIf
            \State \textbf{break}
        \EndIf
        \If{manager\_dict['found\_flag']}
            \State \textbf{break}
        \EndIf
    \EndFor
    \State Print result message
\EndProcedure

\Procedure{Step}{point, scalar, G, public\_key, params}
    \State Define a function to move and update point and scalar
    \State Handle different cases based on x-coordinate of point
    \State \textbf{return} new point and corresponding scalar
\EndProcedure

\Function{GetEccParamsFromServer}{}
    \State Retrieve ECC parameters from server
\EndFunction

\Function{GetPublicKeyFromEntityB}{}
    \State Retrieve public key from Entity B
\EndFunction

\Procedure{PollardsRhoAttackOnEntityB}{params}
    \State Carry out Pollard's rho attack on Entity B in parallel
\EndProcedure

\Procedure{Main}{}
    \State Retrieve ECC parameters
    \State Initialize parameters
    \State Carry out Pollard's rho attack on Entity B
\EndProcedure

\If{name = "main"}
    \State Call Main()
\EndIf
\end{algorithmic}
\end{algorithm}

\begin{figure}[h]
    \centering
    \includegraphics[width=1\linewidth]{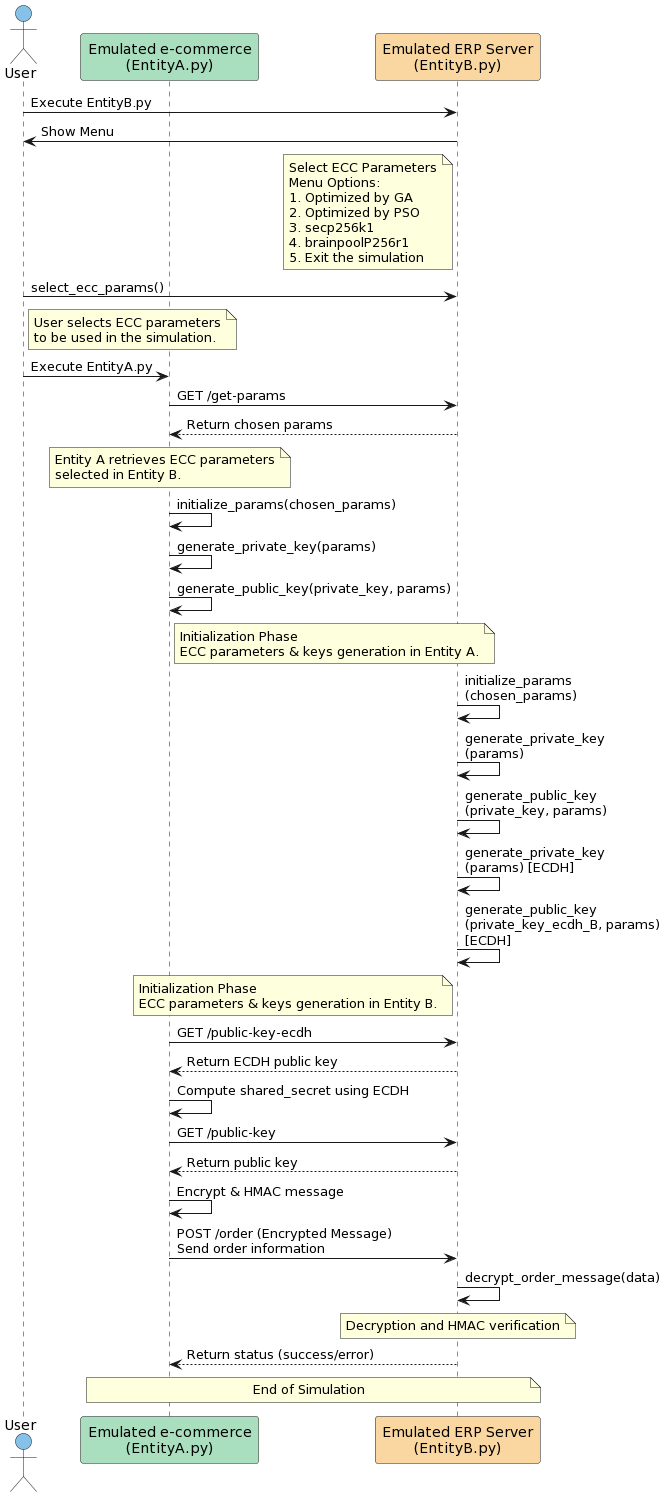}
    \caption{Interaction between \textit{Entity A} (e-commerce) and \textit{Entity B} (ERP server)}
    \label{fig:UML_e-commerce_env}
\end{figure}

The following image is the UML sequence diagram illustrating the interaction between \textit{Entity A} (the emulated e-commerce component) and \textit{Entity B} (the emulated ERP server), as shown in Figure \ref{fig:UML_e-commerce_env}.

\section{Results, Findings, and Analysis}
To summarize our research findings, we divided the results into three stages. First, we ran the GA and PSO artificial intelligence algorithms. Next, we evaluated them within the e-commerce integration simulation. Finally, we compared the results between GA and PSO using the ECC Optimization Criteria detailed in earlier sections of this document.

\subsection{Execution of AI algorithms for ECC optimization}
Each run of the GA and PSO algorithms generates different ECC parameters. This is advantageous for implementation in e-commerce settings or any scenario requiring frequent ECC parameter changes. Despite these differences, fitness function values remain remarkably consistent across runs for both GA and PSO. Below are examples of results from each AI algorithm, as shown in Tables \ref{tab:GAResults} and \ref{tab:PSOResults}:

\begin{table}[h]
\centering
\caption{GA Results}
\label{tab:GAResults}
\begin{tabular}{|l|p{6cm}|}
\hline
Metric & Value \\
\hline
Attack & 0 \\
Min & \(0.0\) \\
Max & \(3.0181340473967544 \times 10^{39}\) \\
Avg & \(3.012097779301965 \times 10^{39}\) \\
Std & \(1.3484001529034777 \times 10^{38}\) \\
\hline
\multicolumn{2}{|l|}{Best Individual Parameters} \\
\hline
Parameter a & \(25947842270905827897659128039154787816323007\)\\
            & \(34210114062670831009467205143790\) \\
Parameter b & \(40136988609592599091657658786458099014083781\)\\
            & \(12405024507582266685792215291693\) \\
Parameter p & \(11572021376927754423644537306382193581670144\)\\
            & \(1977156565361337888165594796740319\) \\
Parameter G & \(0, 72984392299942030530688653046720760764764\)\\
            & \(296696688065665888683496380438139149\) \\
Parameter n & \(11572021376927754423644537306382193581670144\)\\
            & \(1977156565361337888165594796740318\) \\
Parameter h & \(1\) \\
\hline
\end{tabular}
\end{table}

\begin{table}[h]
\centering
\caption{PSO Results}
\label{tab:PSOResults}
\begin{tabular}{|l|p{6cm}|}
\hline
Metric & Value \\
\hline
Attack & 0 \\
Min & \(73.92932346203426\) \\
Max & \(1.5946572521224025 \times 10^{39}\) \\
Avg & \(3.189314504244805 \times 10^{36}\) \\
Std & \(7.1243889397520655 \times 10^{37}\) \\
\hline
\multicolumn{2}{|l|}{Best Particle Parameters} \\   
\hline
Parameter a & \(73916884511138539486074209032992425010519602\)\\
            & \(193355559340498379053138310070272\) \\
Parameter b & \(18466552033214128305449972063322860268226711\)\\
            & \(9339355500319015917133211107328\) \\
Parameter p & \(83920875675429201076002743705901489967077637\)\\
            & \(562817356440692877235699677907597\) \\     
Parameter G & \(2, 52816158108397424543262331025570826905013\)\\
            & \(942926608742347720195343450586800572\) \\
Parameter n & \(11577418207255264997910984803085607312498477\)\\
            & \(0867872005566907726709010164875264\) \\
Parameter h & \(1\) \\
\hline
\end{tabular}
\end{table}

As observed, both GA and PSO produce 256-BIT-based parameters, advantageous for security. However, the fitness function results appear superior in GA \((3.0181340473967544 \times 10^{39})\) compared to PSO \((1.5946572521224025 \times 10^{39})\). The \textit{"Fitness Evolution"} figures emphasize this distinction by showcasing a line graph that tracks the fitness progression over generations or iterations, highlighting the algorithm's convergence. In these graphs, the average fitness value for PSO seems to degrade over time, despite optimizing the Cognitive and Social parameters using grid search. Conversely, GA consistently refines its fitness function, demonstrating a more stable trend.

\begin{figure}[h]
    \centering
    \includegraphics[width=1\linewidth]{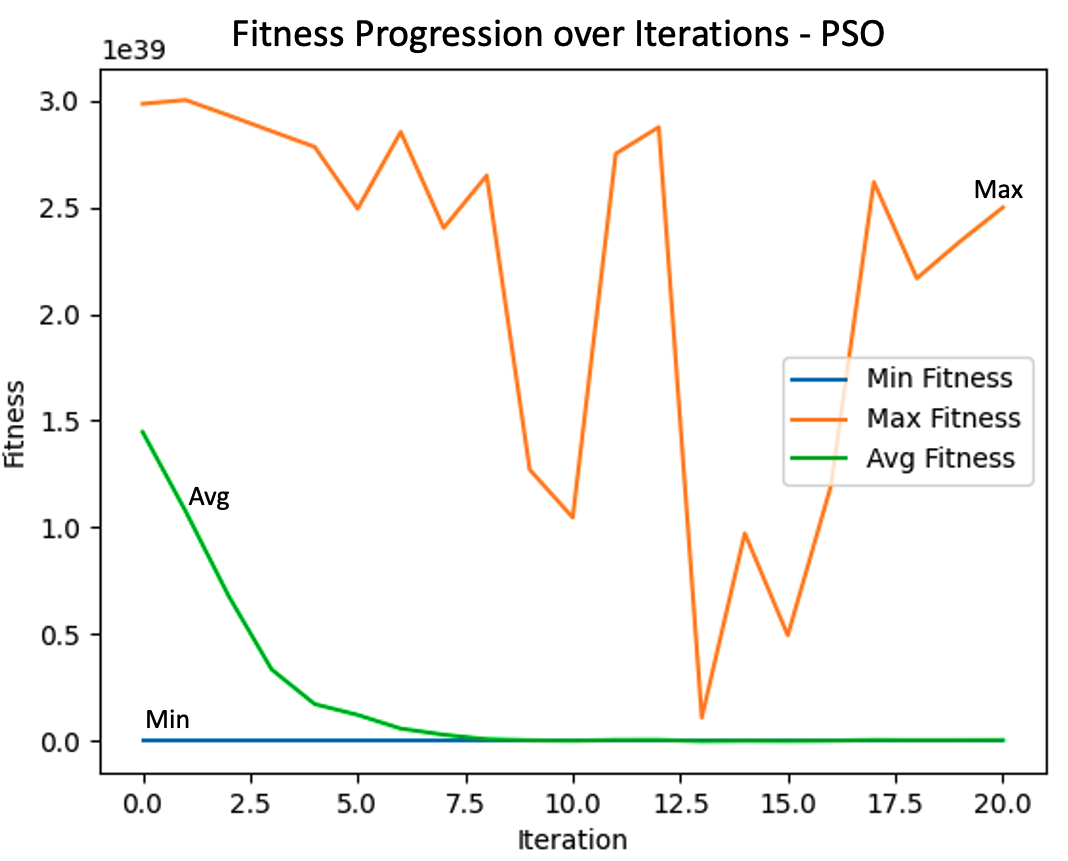}
    \caption{Fitness Progression over Iterations - PSO}
    \label{fig:Fit_Prog_PSO}
\end{figure}

The subsequent charts illustrate these trends in Figures \ref{fig:Fit_Prog_GA} and \ref{fig:Fit_Prog_PSO}:

\begin{figure}[h]
    \centering
    \includegraphics[width=1\linewidth]{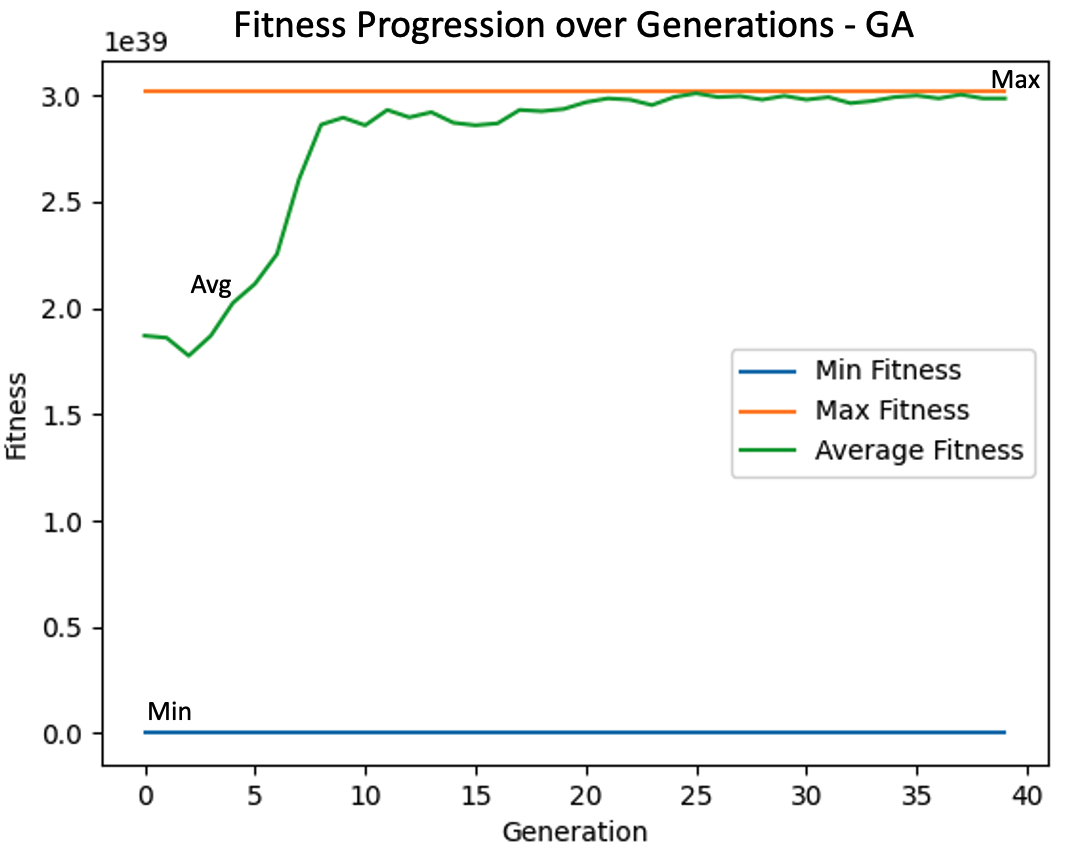}
    \caption{Fitness Progression over Generations - GA}
    \label{fig:Fit_Prog_GA}
\end{figure}

It's worth noting that both algorithms are utilizing the same fitness function. On the other hand, GA generates results much faster than PSO, even though PSO has a feature that allows for early stopping if there's no improvement in the global best fitness for 20 iterations. While GA completes its optimization task in roughly 2 minutes, PSO takes about 22 minutes to finish its task. The tests were conducted on a 2019 15-inch MacBook Pro with the following specifications: Processor: 2.3 GHz 8-Core Intel Core i9; RAM: 16 GB 2400 MHz DDR4; Graphics: Radeon Pro 560X 4 GB and Intel UHD Graphics 630 1536 MB.

\vspace{1em}
\hrule
\hrule
\vspace{1em}

\subsection{Execution of e-commerce integration simulation}
In the simulation testing the transmission of order information in third-party integrations, from the simulated e-commerce to the ERP, the results are promising for both GA and PSO when compared to well-known curves like \textit{secp256k1.txt} and \textit{brainpoolP256r1.txt}. The speed of encrypting and decrypting messages was nearly identical among the four curve types, ranging from fractions of a second to less than 2 seconds. Moreover, a successful Pollard’s Rho attack could not be executed in any scenario, given the computationally intensive task of determining the private key, even when the simulation ran for a week. This approach emphasizes frequent changes in ECC parameters. Compared to well-known curves, GA and PSO offer an advantage by utilizing novel curves without known values for attackers.

\subsection{Comparison based on ECC Optimization Criteria}
The table below displays the winning algorithm based on the acceptance criteria previously outlined in this paper, as shown in Table \ref{tab:comparisonGA_PSO}:

\begin{table}[H]
    \centering
    \caption{Comparison between GA and PSO}
    \label{tab:comparisonGA_PSO}
    \begin{tabularx}{\columnwidth}{|l|l|X|}
        \hline
        \multicolumn{3}{|c|}{\textbf{Efficiency}} \\
        \hline
        \textbf{Criteria} & \textbf{Winning Algorithm} & \textbf{Justification} \\
        \hline
        Performance & GA & Faster generation of ECC parameters. \\
        \hline
        Flexibility & GA & Consistent evolution of the fitness function even when increasing the number of generations. \\
        \hline
        Robustness & GA & Greater tolerance to noise in initial constants definition. \\
        \hline
        Scalability & GA & Remains faster and more efficient even when increasing generations and parallelizing calculations. \\
        \hline
        Comparability & GA & Lower computational complexity. \\
        \hline
        \multicolumn{3}{|c|}{\textbf{Effectiveness}} \\
        \hline
        \textbf{Criteria} & \textbf{Winning Algorithm} & \textbf{Justification} \\
        \hline
        Security & Both GA and PSO & Neither faced successful attacks in the e-commerce integration scenario. However, during parameter generation, PSO experienced more successful attacks than GA when searching for point collisions. \\
        \hline
        Optimality & GA & Consistently closer to expected maximum values and showed improvement over time. \\
        \hline
        Generalization & Both GA and PSO & Quick in both encryption and decryption processes in the simulated scenario, and neither faced successful attacks. \\
        \hline
        Validity & Both GA and PSO & ECC parameters were free from singular or anomalous curves. \\
        \hline
        Practicality & GA & Superior in performance while being compatible with real-world existing systems. \\
        \hline
    \end{tabularx}
\end{table}

Based on the previous table and the information in this section, Genetic Algorithms (GA) outperform Particle Swarm Optimization (PSO) in ECC parameter optimization, not just for third-party e-commerce integrations but also, in terms of efficiency, across any Elliptic Curve Cryptography scenario.

\section{Future Improvements and Known Limitations}
Here are some areas we believe can be improved in the future. These improvements can be considered as potential next steps or future work:

\subsection{Advanced Parameter Tuning}
The current scripts employs grid search for fine-tuning the initial constants. Employing intricate parameter tuning techniques like random search, Bayesian optimization, or metaheuristic algorithms might pinpoint superior parameter values, enhancing the efficacy of the GA and PSO algorithms.

\subsection{Parallelization}
For PSO, the fitness evaluation along with position and velocity updates for individual particles can be executed concurrently, given their independence. Incorporating parallel processing could drastically curtail computational duration, particularly with larger swarms or increased iteration counts.

\subsection{Hybrid Algorithms}
Merging PSO with alternative optimization algorithms can spawn a hybrid model that capitalizes on the strengths of each algorithm. For instance, integrating GA to evolve the swarm while utilizing PSO for refining solutions could augment solution quality and the robustness of the optimization.

\subsection{Exploration of Alternative AI Techniques}
In light of the promising results yielded by the GA and PSO algorithms in this research, it stands to reason that the exploration of alternative artificial intelligence techniques could further optimize ECC parameter generation. Leveraging the same fitness function and utility components established in this study would serve as a foundational bedrock for the integration of techniques such as deep learning, reinforcement learning, or swarm intelligence variations different from PSO, facilitating a seamless transition and a consistent basis for performance evaluation. Such endeavors could potentially unearth novel approaches that are more efficient, secure, and robust, pushing the boundaries of what can be achieved in cryptographic parameter optimization and ensuring a forward momentum in ECC security research.

\subsection{Improved Fitness Function}
The ai\_ecc\_utils.evaluate function, employed as the script's fitness function, ascertains the security of an elliptic curve. This function could undergo enhancement or be supplanted with an alternate fitness function to steer the PSO algorithm search more effectively. For instance, the fitness function might integrate additional security parameters or be tailored to prioritize specific elliptic curve types.

\subsection{Integrating Diverse Cryptographic Threat Evaluations}
In the future, addressing diverse threats to elliptic curves beyond just Pollard's rho attack is crucial. With historical cryptographic vulnerabilities exposed by techniques like the Pohlig-Hellman method and Baby-step Giant-step [45], and the impending rise of quantum computing introducing threats  like Shor's algorithm, modern ECC methods may be at risk [35]. Integrating and evaluating these varied attacks in the fitness function will be essential for fortified defenses.

\subsection{Quantum Computing Implications}

It's crucial to consider the advent of quantum computers and their potential impact on cryptographic algorithms [35]. As quantum computing technology evolves, both GA and PSO algorithms' performance and security measures need re-evaluation to ensure they remain resilient against quantum threats. Additionally, the designed fitness function for this study, which evaluates the security and efficiency of ECC parameters, could be implemented in quantum environments since it is based on optimization through artificial intelligence, allowing for the study of its behavior in such contexts.

\section{Conclusions}
In light of our comprehensive research and systematic evaluation, it is clear that Genetic Algorithms (GA) are more efficient than Particle Swarm Optimization (PSO) in the optimization of Elliptic Curve Cryptography (ECC) parameters. This assertion is based on the adept fitness function we designed and utilized in our research. GA's superiority is evident in third-party e-commerce integrations. Both algorithms are robust, successfully withstanding attacks in e-commerce integration tests. Nevertheless, GA consistently delivers quicker and more reliable performance, integrating seamlessly with real-world systems. This benefit, along with its efficient evolution and rapid ECC parameter generation, underscores GA's dominance in this field. While PSO does offer distinct advantages, its potential is hampered by its relative inefficiency, especially regarding computational speed. Consequently, we strongly advise stakeholders aiming to optimize ECC parameters to prioritize the GA approach.



%

\end{document}